\documentclass[preprint2]{aastex52}
\usepackage{graphicx,natbib}
\bibpunct{(}{)}{;}{a}{}{,}

\newcommand{\figureone}[3]{%
\clearpage
\begin{figure}[tbp]
\begin{center}
\includegraphics[width=88mm]{#1}
\caption{#3}
\label{#2}
\end{center}
\end{figure}
}

\newcommand{\figuretwo}[3]{%
\clearpage
\begin{figure*}[tbp]
\begin{center}
\includegraphics[width=\textwidth]{#1}
\caption{#3}
\label{#2}
\end{center}
\end{figure*}
}

\newcommand\CIV{\mbox{C\,\textsc{iv}}} 
\newcommand\Ha{\mbox{H$\alpha$}}

\newcommand\OVI{\mbox{O\,\textsc{vi}}} 
\newcommand\FeIX{\mbox{Fe\,\textsc{ix}}} 
\newcommand\FeX{\mbox{Fe\,\textsc{x}}} 
\newcommand\FeIXX{\mbox{Fe\,\textsc{ix}/\textsc{x}}} 
\newcommand\FeXII{\mbox{Fe\,\textsc{xii}}}
\newcommand\FeXV{\mbox{Fe\,\textsc{xv}}}
\newcommand\FeXXIV{\mbox{Fe\,\textsc{xxiv}}}

\newcommand\mbxAA{\mbox{\AA}}
\newcommand\km{\ensuremath{\mathrm{km}}}
\newcommand\kms{\ensuremath{\mathrm{km}\,\mathrm{s}^{-1}}}
\newcommand\s{\ensuremath{\mathrm{s}}}
\newcommand\mHz{\ensuremath{\mathrm{mHz}}}
\newcommand\K{\ensuremath{\mathrm{K}}}

\begin{document}

\title{Fourier analysis of active-region plage}
\shorttitle{Fourier analysis of active-region plage}
\author{A.G.~de~Wijn}\altaffilmark{1}
\email{A.G.deWijn@astro.uu.nl}
\and
\author{B.~De~Pontieu}\altaffilmark{2}
\email{bdp@lmsal.com}
\and
\author{R.J.~Rutten}\altaffilmark{1,3}
\email{R.J.Rutten@astro.uu.nl}
\shortauthors{de Wijn, De Pontieu, \& Rutten}

\altaffiltext{1}{Sterrekundig Instituut, Utrecht University, Postbus~80\,000, 3508~TA~Utrecht, The~Netherlands}
\altaffiltext{2}{Lockheed Martin Solar and Astrophysics Laboratory, 3251 Hanover Street, Org.~ADBS, Building 252, Palo Alto, California 94304, USA}
\altaffiltext{3}{Institute of Theoretical Astrophysics, Oslo University, P.O.~Box~1029 Blindern, N-0315 Oslo, Norway}

\begin{abstract}
We study the dynamical interaction of the solar chromosphere with the transition region in mossy and non-mossy active-region plage.
We carefully align image sequences taken with the Transition Region And Coronal Explorer (TRACE) in the ultraviolet passbands around $1550$, $1600$, and $1700~\mbxAA$ and the extreme ultraviolet passbands at $171$ and $195~\mbxAA$.
We compute Fourier phase-difference spectra that are spatially averaged separately over mossy and non-mossy plage to study temporal modulations as a function of temporal frequency.
The $1550$ versus $171$~\AA\ comparison shows zero phase difference in non-mossy plage.
In mossy plage, the phase differences between all UV and EUV passbands show pronounced upward trends with increasing frequency, which abruptly changes into zero phase difference beyond $4$\,--\,$6~\mHz$.
The phase difference between the $171$ and $195$~\AA\ sequences exhibits a shallow dip below $3~\mHz$ and then also turns to zero phase difference beyond this value.
We attribute the various similarities between the UV and EUV diagnostics that are evident in the phase-difference diagrams to the contribution of the \CIV\ resonance lines in the $1550$ and $1600$~\AA\ passbands.
The strong upward trend at the lower frequencies indicates the presence of upward-traveling disturbances.
It points to correspondence between the lower chromosphere and the upper transition region, perhaps by slow-mode magnetosonic disturbances, or by a connection between chromospheric and coronal heating mechanisms.
The transition from this upward trend to zero phase difference at higher frequencies is due to the intermittent obscuration by fibrils that occult the foot points of hot loops, which are bright in the EUV and \CIV\ lines, in oscillatory manner.
\end{abstract}

\keywords{Sun: chromosphere -- Sun: transition region -- Sun: UV radiation}

\maketitle

\section{Introduction}\label{sec:introduction}

In this paper, we compare the dynamics of solar active-region ``moss'' with the dynamics of non-mossy plage.
The context is the dynamical structure of the transition region as interface between the chromosphere and the corona.
The transition region is highly complex due to the fine structure of the chromosphere, which is dominated on small scales by jet-like features.
The latter are observed in \Ha\ as spicules at the limb, and as active-region fibrils and quiet-sun mottles on the disk.
They bring cold ($10^4~\K$), dense plasma up to greater heights than expected from hydrostatic equilibrium
	\citep[e.g.,][]{1995ApJ...450..411S}. 
Understanding the formation and behavior of these structures is of key importance, because their role as agents in the momentum and energy balances of the outer solar atmosphere remains unclear.

Moss is characterized by highly dynamic dark inclusions over a bright background in active-region plages as observed in image sequences taken through the EUV passbands of the Transition Region and Coronal Explorer
	\citep[TRACE,][]{1999SoPh..187..229H}. 
It is best visible in images taken in the \FeIXX\ passband around $171~\mbxAA$, but it is also present in images taken in the other EUV passbands around $195$ (\FeXII) and $284~\mbxAA$ (\FeXV).
The bright background consists of $\sim10^6~\K$ foot points of hot, high-pressure coronal loops at heights low enough to be obscured by cool chromospheric jets
	\citep{1999SoPh..190..409B, 
	1999SoPh..190..419D, 
	1999ApJ...520L.135F}. 
Non-mossy plage does not have such overlying high-density coronal loops, so that its emission in the TRACE EUV passbands is of lower intensity and formed over a large height range.

Previous studies by
	\cite{2003ApJ...590..502D} 
	and
	\cite{dewijn+depontieu2006}
concentrated on the spatial correlations between various chromospheric and transition-region diagnostics in plage.
	\cite{2003ApJ...595L..63D} 
studied individual positions in plage with wavelet techniques, and found that oscillations in moss correspond well with oscillations in \Ha\ diagnostics, and occasionally also with oscillations in photospheric Dopplergrams.
In the present analysis, we concentrate on temporal brightness modulations in image sequences of mossy and non-mossy plage taken in TRACE's EUV $171$ and $195$~\AA\ passbands and its UV passbands around $1550$, $1600$, and $1700~\mbxAA$.

\section{Observations and Data Reduction}\label{sec:observations}

We analyzed image sequences recorded simultaneously on 1999 June~4 between 8:31 and 9:43~UT and on 1998 July~1 between 19:36 and 20:53~UT.
The target of the 1999 observations was plage around NOAA active region 8558 at a heliocentric viewing angle of $20$~degrees ($\mu=0.94$).
TRACE took images sequentially in the three UV passbands at $1550$, $1600$, and $1700~\mbxAA$, and in the \FeIXX\ EUV passband at $171~\mbxAA$, at a cadence of $42\pm9~\s$.
The 1998 observations targeted plage around NOAA active region 8235 at a heliocentric viewing angle of $40$~degrees ($\mu=0.77$) in the $1600$~\AA\ UV passband and in the EUV $171$~\AA\ (\FeIXX) and $195$~\AA\ (\FeXII) passbands.
The size of all images is $512\times512$~pixels, and their angular resolution is $0.5\arcsec$ per pixel.
The UV passbands sample the low chromosphere in the respective continua, but the $1550$ and $1600$~\AA\ passbands also have a transition-region contribution from \CIV\ lines.

The images were corrected for dark current and were flat-fielded by the SolarSoft procedure \texttt{trace\_prep} described in the TRACE Analysis Guide\footnote{\url{http://moat.nascom.nasa.gov/~bentley/guides/tag}}.
Cosmic ray hits were removed iteratively from each image with a median filter that removes spikes exceeding a $20\%$ threshold and replaces them with the median of the surrounding pixels until less than $1000$ particle hits are detected.
The UV sequences were subsequently aligned with high precision using the Fourier cross-correlation technique implemented in the SolarSoft routine \texttt{tr\_get\_disp} in three steps following the procedure developed by
	\cite{2005A&A...430.1119D}.

The EUV sequences of the 1999 observations were each first aligned frame by frame.
Next, the average displacement to the $1600$~\AA\ UV passband was calculated by comparing the position of the light bridge of the large sunspot around $(170\arcsec,275\arcsec)$.
This light bridge has a cross-like shape in the $1600$~\AA\ image.
The corona above it is hot and of high-pressure, so that the $171$~\AA\ emission is formed at low altitude, and closely matches the $1600$~\AA\ emission in shape.
This allows for accurate co-alignment of the UV and EUV passbands.
The final sequence was resampled in a single interpolation step from the original image sequence.

There also are several white-light images in the 1999 observations.
We use these to identify the positions of the sunspots and pores in the field of view.
We align each of the white-light images to the logarithm of the brightness of the $1600$~\AA\ image nearest in time.

The 1998 sequences are similar, except for the absence of $1550$ and $1700$~\AA\ UV sequences, and the addition of a $195$~\AA\ sequence.
We process them akin to the 1999 sequences, simplifying the procedure of
	\cite{2005A&A...430.1119D}
to align the images of the $1600$~\AA\ sequence, and treating the $195$~\AA\ sequence the same as the $171$~\AA\ sequence.
In this case, the average offset to the $1600$~\AA\ sequence was computed by alignment on the central plage around $(550\arcsec,240\arcsec)$ (cf.~Fig.~\ref{fig:sample1998}).

\placefigure{fig:sample1999}

Figure~\ref{fig:sample1999} displays a sample set of images from the 1999 observations in the $1600$ and $171$~\AA\ passbands, together with a white-light image for context information.
The $1550$ and $1600$~\AA\ passbands include the \CIV\ lines at $1548$ and $1550~\mbxAA$ that are formed at temperatures around $60$\,--\,$250\times10^3~\K$, and sample the transition region.
There are also considerable contributions from the surrounding continuum due to the broad passbands ($20$ and $275~\mbxAA$, respectively).
The $1700$~\AA\ passband has a width of $200~\mbxAA$, but does not contain the \CIV\ lines.
The $171$~\AA\ passband samples emission from the transition region and low corona primarily from \FeIX\ and \FeX\ ions at temperatures of about $10^6\,\mathrm{K}$.

\placefigure{fig:sample1998}

Figure~\ref{fig:sample1998} displays a similar sample set of images from the 1998 observations in the $1600$, $171$ and $195$~\AA\ passbands.
The $195$~\AA\ passband samples emission primarily from \FeXII\ ions at temperatures of about $10^6~\K$ and \FeXXIV\ ions under flaring conditions.
It is similar to the $171$~\AA\ passband, sampling much the same structures in the transition region and low corona.

We selected mossy and non-mossy plage in the field of view for analysis.
We applied a suitable high-pass brightness threshold to the time-averaged $171$~\AA\ sequence, and then selected those areas that contain moss to produce the mask.
For the non-mossy plage, we select areas from a mask created by applying a high-pass brightness threshold to the time-averaged $1600~\mbxAA$ sequence and a low-pass threshold to the average of the $171$~\AA\ sequence.
Some non-mossy plage in the 1999 observations is not usable, because it is covered by loops in the $171$~\AA\ sequence.
There is no usable non-mossy plage in the 1998 observations.
The resulting masks are smoothed with $10\times10$-pixel boxcar averaging and restricted to those pixels that do not suffer from partial sampling due to solar rotation and image motions.
In case of the 1999 observations, we also exclude those pixels that are insufficiently bright in the average white-light image in order to mask off the sunspots and pores.
The resulting mossy-plage mask for the 1999 observations contains seven patches with a total surface area of $13455$~pixels, while the mask for the 1998 observations consists of a single patch of $7436$~pixels.
The 1999 non-moss mask has $10230$~pixels in three patches.
Figure~\ref{fig:sample1999} shows the outlines of masks of mossy (black) and of non-mossy plage (white), together with the sunspot mask indicated by dashed black contours.
Figure~\ref{fig:sample1998} shows the mossy-plage mask outlined in black.

\section{Analysis and Results}\label{sec:analysis}

We computed Fourier power, coherence, and phase-difference spectra averaged over pixels in mossy and in non-mossy plage using the masks shown in Figs.~\ref{fig:sample1999} and~\ref{fig:sample1998}.
Following
	\cite{2001A&A...379.1052K} 
and
	\cite{2005A&A...430.1119D}, 
the power and coherence were averaged directly over all relevant pixels.
We calculated the coherence per pixel using frequency smoothing over 5 bins.

In the evaluation of phase differences, we employed averaging of the Fourier cross-power vectors as introduced by
	\citet{1979ApJ...229.1147A} 
and
	\citet{1979ApJ...231..570L}. 
This method avoids the problems that arise in averaging due to the $2\pi$ ambiguity, and in addition weighs the samples with their cross-power amplitude.
Furthermore, in the case of pure noise, the summation of the cross-power vectors resembles a random walk around the origin, thus producing a random phase.
As long as the phase-difference spectrum does not display random variation from one frequency to the next, the input contains a signal, be it solar, instrumental, or the result of data processing.
A more recent analysis by
	\citet{2006A&A...452.1059O} 
also follows this method of computing phase differences.

\placefigure{fig:confusomaximus}

The Fourier spectra resulting from the 1999 sequences are shown in Fig.~\ref{fig:confusomaximus}.
The power, coherence, and phase-difference spectra in the leftmost two columns are similar to those of
	\cite{2001A&A...379.1052K} 
and
	\cite{2005A&A...430.1119D}, 
but are less accurate, because of the smaller number of pixels transmitted by the mask and the shorter sequence length.
Note that mossy and non-mossy plage exhibit similar behavior.
The phase differences are well-defined and small.

The $171$~\AA\ passband samples much hotter structures of the atmosphere.
In non-mossy plage the $1700$\,--\,$171$~\AA\ and $1600$\,--\,$171$~\AA\ phase-difference spectra are very noisy, which is indicative of the lack of any correspondence.
However, the $1550$\,--\,$171$~\AA\ phase-difference spectrum is nearly constant at zero degrees up to high frequency, about $8~\mHz$.

The mossy plage exhibits radically different behavior.
All three UV sequences show a well-defined phase difference with the $171$~\AA\ sequence, with a strong upward trend up to about $4~\mHz$.
The $1600$ and $1550$\,--\,$171$~\AA\ phase-difference spectra show a flat tail at $0$~degrees beyond $4~\mHz$.
The $1700$~\AA\ sequence (third column) wraps around a second time, while the $1600$~\AA\ sequence (fourth column) reaches a higher phase difference than the $1550$~\AA\ sequence (fifth column).
The frequency at which the phase difference ceases the upward trend and falls to zero appears to shift somewhat towards lower frequencies.
The flat tail at zero degree is most pronounced in the $1550$\,--\,$171$~\AA\ panel (fifth column).
It appears to be very noisy in the $1700$\,--\,$171$~\AA\ panel (third column).

The power spectra are normalized by their value in the second frequency bin, so they can only be compared relatively.
The mossy-plage power spectra have a lower relative noise floor on account of the larger brightness variation.
In the UV passbands, the power hump due to $5$-minute oscillations is smaller in mossy plage than in non-mossy plage.
It is absent in the $171$~\AA\ power spectra for either network or mossy plage.

The coherence spectra show large values at the lowest and highest frequencies, which however is an artifact resulting from the 5-bin frequency smoothing.
The coherence is small for the UV versus EUV comparisons, which indicates that the signals in the phase-difference spectra are weak, as is shown by the large scatter in the grayscale background of the phase-difference panels.
In contrast, in the first two columns the strong signal has coherence well above the noise floor, and small grayscale scatter around the average value.

\placefigure{fig:confuse1998}

Figure~\ref{fig:confuse1998} shows the results for the 1998 sequences.
The upward trend is present in both the $1600$\,--\,$171$~\AA\ and the $1600$\,--\,$195$~\AA\ phase-difference spectra, up to about $6~\mHz$.
The $171$\,--\,$195$~\AA\ phase-difference spectrum shows a slightly negative phase difference up to $3~\mHz$ with reasonable coherence, and a constant zero-degree phase difference beyond.
The power spectra show behavior similar to those in Fig.~\ref{fig:confusomaximus}.

\section{Discussion}\label{sec:discussion}

\subsection{Non-mossy Plage}

The well-defined $1550$\,--\,$171$~\AA\ phase difference for non-mossy plage in the top-rightmost panel of Fig.~\ref{fig:confusomaximus} shows that there is common modulation in the $1550$ and the $171$~\AA\ sequences.
The correspondence is weak, as is evident from the extensive scatter and the low coherence, but nevertheless significant.
The $1600$ and $1700$\,--\,$171$~\AA\ panels in the top row show noise, indicating that this apparent signal is not caused by low brightness, cadence irregularity, or non-simultaneous sampling.
We attribute the larger noise in these panels to the smaller or absent contribution of the \CIV\ lines, which sample some of the structures and dynamics governing the $171$~\AA\ signal.
The $1600$~\AA\ panel shows somewhat less noise than the $1700$~\AA\ one, as is expected from the small contribution of \CIV\ in case of the former, and its absence in the latter.
It does not seem unreasonable to assume there is some signal present in the $1600$\,--\,$171$~\AA\ phase-difference spectrum, but drowned in noise due to the reduced \CIV\ contribution.

We cannot establish from these observations what the cause is of the apparent dynamical correspondence between the $171$ and $1550$~\AA\ brightness in non-mossy plage.
Small-scale energy deposition by, e.g., wave energy deposition, reconnection, or turbulent heating may cause simultaneous brightening.
On the other hand, the $171$~\AA\ passband also contains \OVI\ lines, which sample temperatures similar to the \CIV\ lines.
Alternatively, coronal rain in overlying loops may cause intermittent brightening in both diagnostics
	\citep[cf.][]{2005A&A...436.1067M}.
Despite our effort to select only those regions that do not suffer much from $171$~\AA\ loops, avoiding them completely is impossible.
A study of loop areas without underlying plage shows random phase difference, suggesting that they are not to blame.
However, the statistics are poor as a result of a limited number of pixels in the loop mask.

Summarizing, it is clear that the weak signal is somehow related to the \CIV\ lines sampled best by the $1550$~\AA\ passband, and less so by the $1600$~\AA\ one.
A number of processes may contribute to the signal, and we cannot exclude with certainty that it is caused by overlying loops rather than by the plage.

\subsection{Low-frequency Behavior of Mossy Plage}

There is a strong upward trend in all UV\,--\,EUV phase-difference spectra at low frequencies.
It continues up to higher frequency in the $1600$\,--\,$171$~\AA\ phase-difference spectrum than in the $1550$\,--\,$171$~\AA\ spectrum, and yet further in the $1700$\,--\,$171$~\AA\ spectrum.
The upward trend extends to frequencies up to $4~\mHz$ in the 1999 sequences, and somewhat further in the 1998 sequences.
The phase differences depend linearly on frequency, suggesting that disturbances propagating upward through strong flux tubes travel a fixed distance with a velocity independent of frequency.
It wraps around to zero at around $7$-minute periods, indicating a delay of approximately $400~\s$ between the lower and higher layers.
In principle, one would expect the slope of the trend to change for spectra of different UV\,--\,EUV pairs, because of the height difference between the layers sampled by the passbands.
However, the linear relation appears the same within experimental errors in all UV\,--\,EUV phase-difference spectra, irrespectively of the specific UV or EUV passband.
We attribute this to the large height difference between the layers sampled by UV and the EUV passbands.
Pairs of UV or EUV passbands sample layers with comparatively small height difference, as indicated by a phase difference close to zero degrees (Fig.~\ref{fig:confusomaximus} first and second column, and Fig.~\ref{fig:confuse1998} third column), so that the height difference and associated travel time do not depend much on the choice of UV or EUV passband.

What is the upward trend caused by?
The trend is observed in all UV passbands, and it is therefore not related to \CIV\ emission, which is present only in the $1550$ and $1600$~\AA\ passbands.
Also, it is not seen in the non-mossy panels, indicating that it is unlikely to be an artifact of instrumental origin or related to data processing.
It thus seems of a solar origin: brightness changes in the UV passbands are followed about $400~\s$ later by changes in the emission in the EUV passbands.

The UV passbands are dominated by emission from low chromospheric heights.
Plage is brighter than its surroundings in these passbands.
This excess brightness is related to enhanced chromospheric heating in areas of strong magnetic flux.
The EUV passbands in mossy plage are dominated by emission from $1$\,--\,$1.5\times10^6~\K$ plasma in a thin layer at the transition-region foot points of hot ($>3$\,--\,$4\times10^6~\K$) coronal loops.
	\cite{1999SoPh..190..409B}, 
	\cite{1999ApJ...520L.135F}, and 
	\cite{2000ApJ...537..471M} 
showed that moss is heated by thermal conduction from the hot loops above the moss.
Our observations indicate that brightness changes on timescales of $5$\,--\,$30$~minutes in the low chromosphere are followed by brightness changes on similar timescales in the upper transition region about $400~\s$ later.
This relationship has not been reported before and may offer an intriguing clue to the nature of chromospheric and coronal heating, and how they are related.

What is the cause of this relationship?
It is difficult to imagine that real wave packets with such long periods are present and propagating.
Changes on timescales of order $10$\,--\,$30$~minutes are typically considered to be dominated by solar evolution, i.e., slow variations of the atmosphere because of, e.g., evolution of the magnetic network or plage as a result of buffeting by convective flows, reconfiguration or emergence of magnetic field, etc.
Disturbances associated with such slow changes may propagate up in the atmosphere, while real wave packets would find it difficult to maintain their oscillatory identity.
It would seem that our observations show solar evolution causing chromospheric brightness changes that are followed by an associated modulation in upper transition-region brightness.

The fixed time delay of about $400~\s$ may provide an important clue to the nature of the correlation between the chromosphere and the upper transition region.
The UV passbands have formation heights of a few hundred km above the photosphere
	\citep{2005ApJ...625..556F}. 
The EUV moss is estimate to be formed several thousand km above the photosphere, over a range of heights between $2000$ and $5000~\km$
	\citep[see, e.g.,][]{2000ApJ...537..471M}, 
since it is buffeted up and down by dynamic fibrils that shoot up from the chromosphere.
A height difference between UV and EUV passband formation heights of order $1500$\,--\,$4500~\km$ would lead to propagation speeds between $3.5$ and $11~\kms$, assuming a vertical wave guide.
These velocities are too low for slow-mode magnetosonic disturbances if the moss occurs at $2000~\km$, and low, but perhaps possible if the moss occurs at $4500~\km$.
While the magnetic field is likely mostly vertical in the interior of plage, it can be strongly inclined at the edges.
However, for our pixel-based analysis, the large height difference and the size of the TRACE pixels limits the angle to about $10$~degrees.
The maximum inclinations are then $\sim30$ and $\sim50$~degrees for the 1999 and 1998 observations, respectively, assuming that the offset introduced by the viewing angle is not (partly) corrected by the alignment procedure.
For the 1999 observations, these corrections could increase the estimated propagation speeds to more reasonable values between $6.5$ and $16~\kms$.
The correction is small ($\sim15\%$) for the 1998 observations.
Is it possible for slow-mode disturbances with long periods of $5$\,--\,$30$~minutes to actually propagate upward?
While their periods are longer than the acoustic cutoff, it is possible for the initial disturbance, i.e., before any evanescent wave pattern is established, to propagate.
So, perhaps the density and temperature perturbations associated with slow-mode disturbances, maybe driven by granulation, supergranulation, waves, or reconnection, could lead to chromospheric brightness changes, and upper transition-region brightness changes $400~\s$ later.
Note, however, that the time delay of $7$~minutes is quite long for slow-mode disturbances.
This problem is even more severe for fast magneto-acoustic disturbances.

Perhaps then the relationship between chromospheric and upper transition-region brightness is not mediated by waves, but indicative of a connection between chromospheric and coronal heating mechanisms?
In such a scenario, heating in the chromosphere (e.g., through dissipation of currents) would be followed by heating in the corona, which, some time later, would lead to increased emission in the transition-region moss, since the latter is heated by thermal conduction from hot loops above.
In this case, the time delay would be determined by a combination of the delay between chromospheric and coronal heating events, and the timescale associated with moss brightness changes.
The brightness of moss emission can change because of a change in temperature (related to thermal conduction), and because of a change in density (related to chromospheric evaporation or coronal loop dynamics).
It is difficult to estimate these timescales without advanced numerical models.
Similarly, the delay between chromospheric and coronal heating events is unknown, and can only be addressed by advanced radiative magnetohydrodynamics simulations of the low atmosphere that include the corona.
Our findings can thus provide an interesting observational constraint on chromospheric and coronal heating models.

\subsection{High-frequency Behavior of Mossy Plage}

The phase difference of the $1600$ or $1550$~\AA\ sequence in relation to the $171$~\AA\ one is close to zero for frequencies exceeding $5~\mHz$ in the 1999 June~4 sequences.
The 1998 sequences have more noise due to a reduced number of pixels in the mossy-plage mask, but the same behavior appears present in the $1600$\,--\,$171$~\AA\ phase-difference spectrum between $6$~and $12~\mHz$.
Intensity changes in moss are known to be directly related to chromospheric fibrils (as observed in \Ha) that have lifetimes around $5$~minutes
	\citep{2003ApJ...595L..63D}. 
We therefore attribute this behavior to active-region fibrils that periodically occult the bright foot points of hot loops.
The ``fibrils shutter'' causes simultaneous extinction of \FeIXX\ and \CIV\ emission, thus producing a signal with zero phase difference.
It has Fourier power up to high frequencies due to its instantaneous blocking of bright foot points, causing step-like intensity modulation.
The $1600$~\AA\ passband includes the \CIV\ lines, but with more surrounding continuum than the $1550$~\AA\ passband.
It is affected by the shutter, as is indicated by the turn to zero-degrees phase difference, but not as much as the $1550$~\AA\ sequence.
The $1600$\,--\,$195$~\AA\ phase-difference spectrum (Fig.~\ref{fig:confuse1998} second column) is expected to behave in a similar fashion.
There is too much noise, however, to confirm this, most likely because moss becomes less pronounced in the EUV passbands at longer wavelengths.

The fibril shutter does not act on the $1700$~\AA\ sequence, since it does not include the \CIV\ lines.
Accordingly, it does not show a clear zero-degree phase difference.
It appears that the $1700$\,--\,$171$~\AA\ phase-difference spectrum does not continue its upward trend to the Nyquist frequency of our observations, but becomes noisy and flattens out somewhat around $6$\,--\,$8~\mHz$.
Many non-linear effects may disrupt the process that causes the trend at higher frequencies.
There is no evolution on such short time scales, so that the trend may disappear if it is caused by evolutionary processes.

\subsection{Model of Mossy Plage}

\placefigure{fig:cartoon}
\placefigure{fig:phasesim}

Figure~\ref{fig:cartoon} shows a cartoon of mossy plage.
Fibrils intermittently obscure \CIV, \FeIXX, and \FeXII\ emission, but do not affect emission from the UV continuum.
A signal with a broad signature in Fourier space at low frequencies propagates upward.
The observer thus measures the signal propagating from low layers to higher ones while it is modulated with an external fibril shutter.
The timing of the shutter is identical in all passbands, but with varying efficiency.

One can use a toy model to confirm that an upward propagating signal with a superimposed shutter reproduces the phase-difference spectrum observed in mossy plage.
Artificial observations were constructed with identical exponential power spectra, but the upper layer was retarded by $160$~degrees per~mHz ($444$~s).
The shutter is simulated by $150$-s wide block functions with a period of $300\pm10~\s$.
The reduced effect of the shutter on the UV passbands was simulated by applying $10\%$ shutter to the intensity of the lower level, and $50\%$ to the upper level.
We so created 200 independent samples, and computed an average phase-difference spectrum from them.
The result is represented by the solid curve in Fig.~\ref{fig:phasesim}.
It exhibits the desired behavior, i.e., a strong upward trend at low frequencies, and a small phase difference at high ones.
The dotted curve displays the phase-difference spectrum if the shutter is not installed.
It shows the linear dependence on frequency that was inserted into the artificial observations.
The point at which the trend turns toward zero phase difference can be adjusted toward lower frequency by decreasing the period of the shutter, or by increasing the spread in the period, the efficiency of the shutter, or the slope of the power spectrum.

\subsection{Mossy Plage in EUV Passbands}

The observed $171$\,--\,$195$~\AA\ phase-difference spectrum is reassuring in that it indeed shows near-zero degrees up to the Nyquist frequency, with little noise up to $6~\mHz$.
The shutter causes a phase difference of zero degrees, starting at as little as $3~\mHz$ and continuing up to the Nyquist frequency.
	\cite{2003ApJ...590..502D} 
similarly found good correlation and zero second lag time between the $171$ and $195$~\AA\ passbands.
The increased noise upward from $6~\mHz$ is most likely due to the reduced clarity of moss in the $195$~\AA\ passband as compared to the $171$~\AA\ one.
The spectrum has a dip around $1~\mHz$, similar to the $1700$\,--\,$1600$~\AA\ and $1700$\,--\,$1550$~\AA\ phase-differences spectra (Fig.~\ref{fig:confusomaximus} first and second column).
The latter is caused by reversed granulation
	\citep{2005A&A...431..687L}. 
However, there is no reversed granulation visible in the $171$ or $195$~\AA\ passband.
It is tempting to assume that this dip is evidence of loop cooling.
This would cause a slight delay between corresponding passbands, with emission in $195~\mbxAA$ preceding $171~\mbxAA$.
It reaches a minimum of about $-7.5$~degrees at $0.65~\mHz$, corresponding with a time delay of $75~\s$.
	\cite{2003ApJ...593.1164W} 
find longer cooling times.
However, the loops they studied are less dense and colder than those that originate from mossy plage, that are expected to have a steeper gradient and therefore have a shorter cooling time.

\section{Summary and Conclusion}\label{sec:conclusion}

We have compared phase-difference spectra between the TRACE UV passbands around $1550$, $1600$, and $1700~\mbxAA$ and the EUV passbands around $171$ and $195~\mbxAA$ in non-mossy and mossy plage.

We find a nearly constant zero phase difference between the $1550$ and $171$~\AA\ passbands in non-mossy plage.
We attribute this to \CIV\ emission in the $1550$~\AA\ passband that corresponds with emission in the $171$~\AA\ passband, possibly from small-scale energy deposition events, from \OVI\ contamination in the $171$~\AA\ passband, or from coronal rain in loops overlying the non-mossy plage.

The UV\,--\,EUV phase-difference spectra in mossy plage show a strong upward trend with increasing frequency from zero up to $4$\,--\,$6~\mHz$.
Beyond $4$\,--\,$6~\mHz$, they show a constant zero phase difference.
In the $171$\,--\,$1700$~\AA\ phase-difference spectrum, the upward trend continues to higher frequencies than in the $171$\,--\,$1600$~\AA\ spectrum.
It continues yet further in the $171$\,--\,$1550$~\AA\ spectrum.
Its absence in the non-mossy panels indicates that it is not an artifact of instrumental origin, or due to data processing.

A possible cause of the upward trend in the phase-difference spectrum might be propagation of slow-mode magnetosonic disturbances.
However, the delay of $400~\s$ sets a low average propagation speed.
Alternatively, it may be that chromospheric and coronal heating may be somehow correlated, so that heating in the chromosphere is followed by heating in the upper transition region through thermal conduction from above by excess coronal heating.
Such a connection between chromospheric and coronal heating is potentially very interesting for heating models.

We attribute the zero phase difference at higher frequencies in mossy plage to active-region fibrils occulting the foot points of hot loops, causing moss in the EUV passbands, and also causing simultaneous extinction of the \CIV\ lines in the $1600$ and $1550$~\AA\ passbands.
The $1700$~\AA\ passband does not include the \CIV\ lines, and is therefore not affected by the shutter-like action of the fibrils.
A simple toy model reproduces this behavior well.

The $171$\,--\,$195$~\AA\ phase-difference spectrum of mossy plage shows a shallow dip up to about $3~\mHz$.
We speculate that this is caused by loop cooling.
The phase difference is zero beyond $3~\mHz$, which is again a result of the fibril shutter causing simultaneous extinction in both passbands.

\acknowledgments
AdW acknowledges travel support from the Leids Kerkhoven-Bosscha Fonds and hospitality at LMSAL and MSU.
BDP was supported by NASA grants NAG5-11917, NNG04-GC08G, and NAS5-38099 (TRACE).



\figuretwo{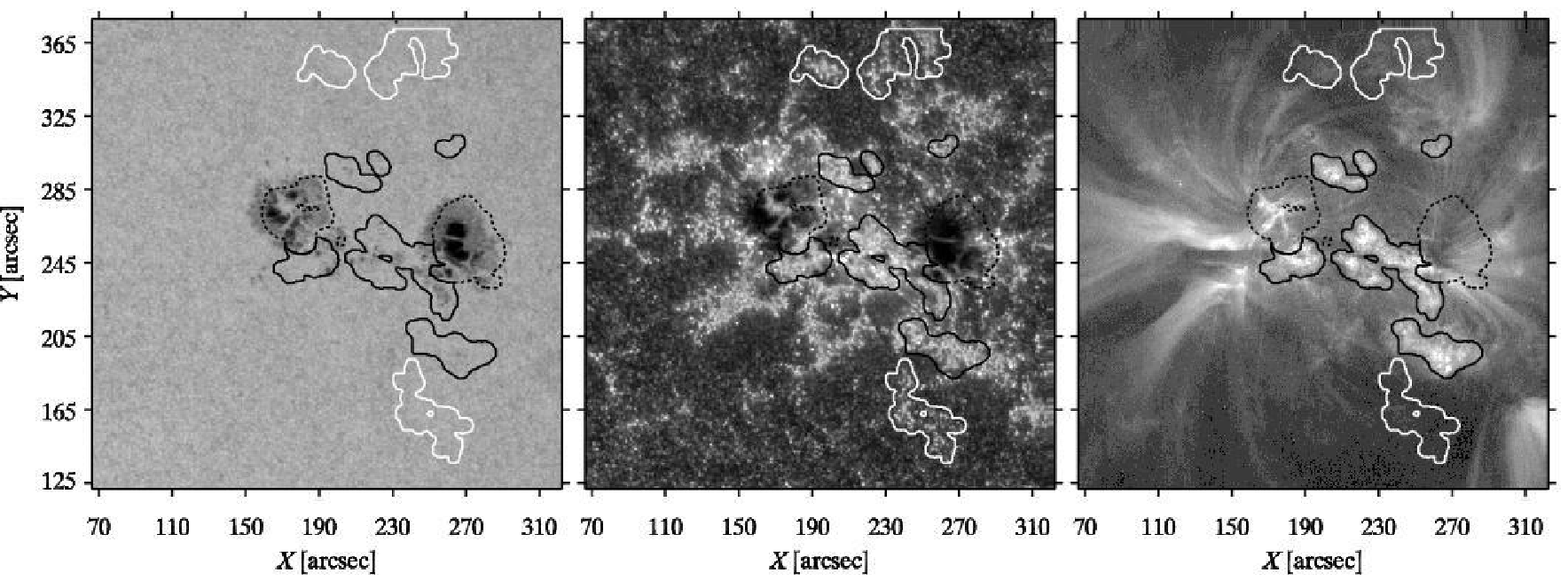}{fig:sample1999}{%
A sample set of TRACE images from the 1999 June~4 sequence.
\emph{Left:} white light, shown here for context information.
\emph{Center:} $1600$~\AA\ UV continuum.
\emph{Right:} $171$~\AA\ EUV \FeIXX.
The images were corrected for dark current and flat-fielded.
The intensities of the $1600$ and $171$~\AA\ images are scaled by taking the square root for better representation of dark structures.
Solar north is up in all TRACE images.
Areas of mossy and non-mossy plage are surrounded by black and white contours, respectively.
The locations of the sunspots and pores are indicated by dashed black contours.
}

\figuretwo{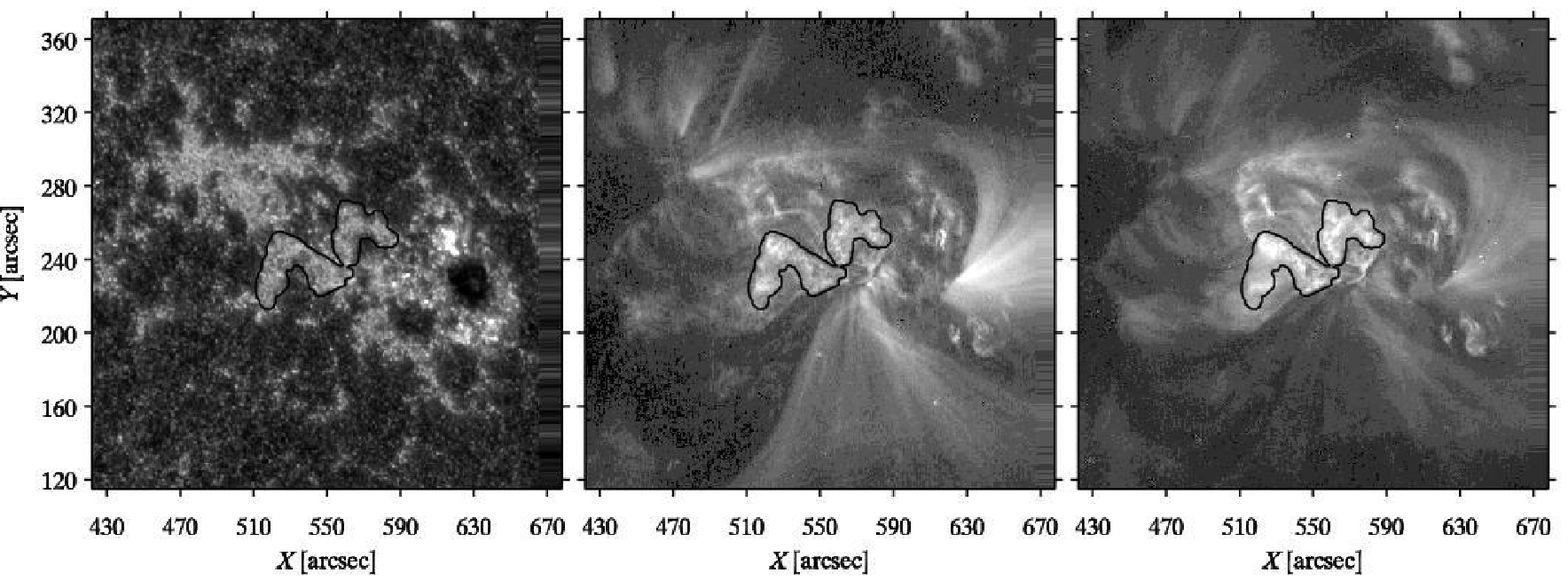}{fig:sample1998}{%
A sample set of TRACE images from the 1998 July~1 sequence.
\emph{Left:} $1600$~\AA\ UV continuum.
\emph{Center:} $171$~\AA\ EUV \FeIXX.
\emph{Right:} $195$~\AA\ EUV \FeXII.
The images were corrected for dark current and flat-fielded.
The intensities of the images are scaled by taking the square root for better representation of dark structures.
Solar north is up in all images.
The area of mossy plage is surrounded by a black contour.
}

\figuretwo{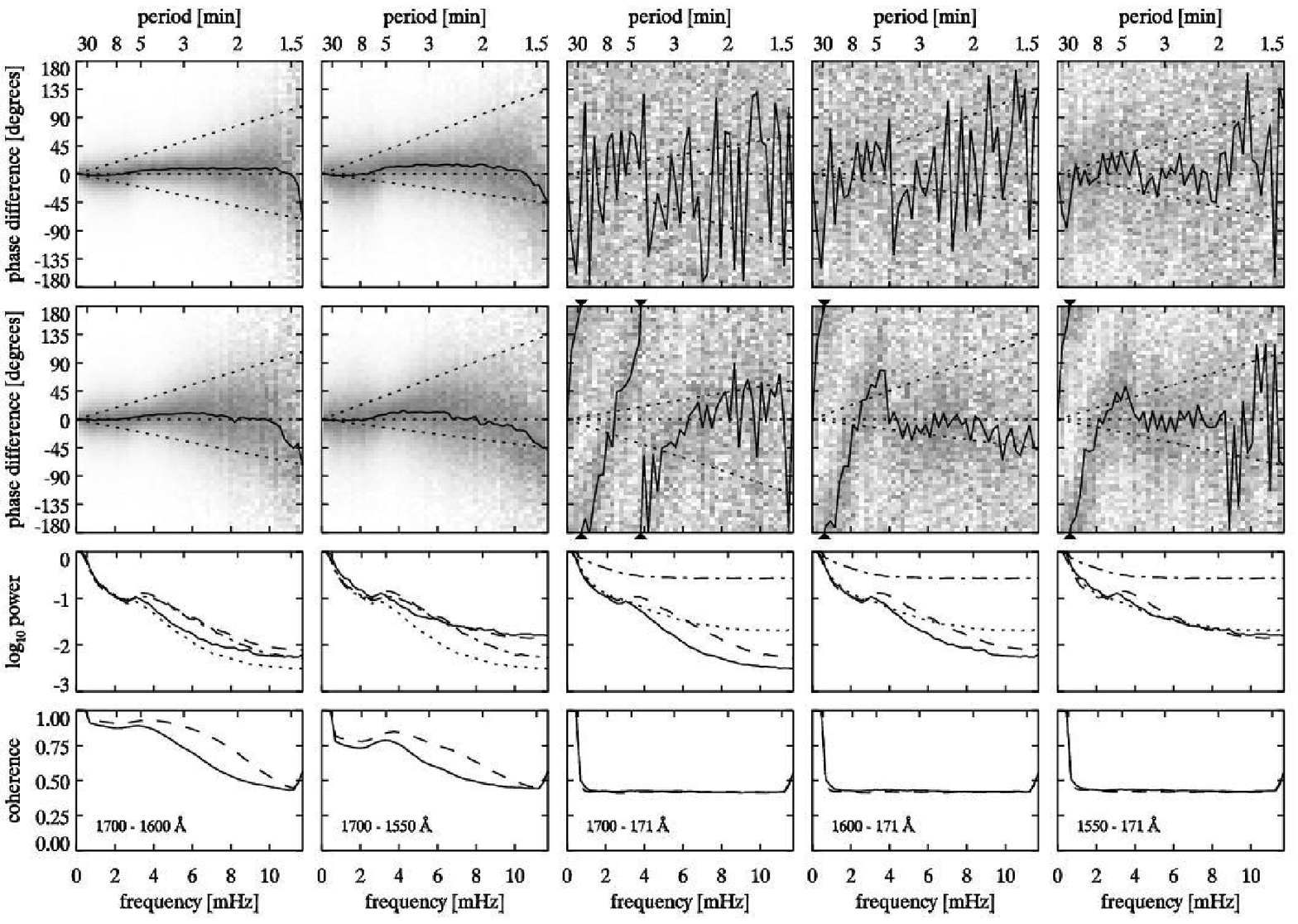}{fig:confusomaximus}{%
Fourier phase-difference and power spectra for non-mossy and mossy plage in the 1999 June~4 sequences for various passbands as specified in the bottom panels of each column.
\emph{First and second row:} phase-difference spectra for non-mossy and mossy plage, respectively.
The dotted lines indicate zero and the phase differences corresponding to the timing offset between the passband samplings.
The solid curves represent weighted average phase differences, and the grayscaled clouds the phase difference of the individual pixels.
Note the $360$-degree phase jumps for mossy plage (second row) in the rightmost three columns as indicated.
\emph{Third row:} corresponding power spectra, each scaled by the power in its second frequency bin.
The solid and dotted curves represent the longer and shorter wavelengths for mossy plage, and the dashed and dashed-dotted curves for non-mossy plage.
\emph{Bottom row:} coherence spectra, solid for mossy, dashed for non-mossy plage.
The noise level is set at $0.4$ by the number of frequency bins in the averaging.}

\figureone{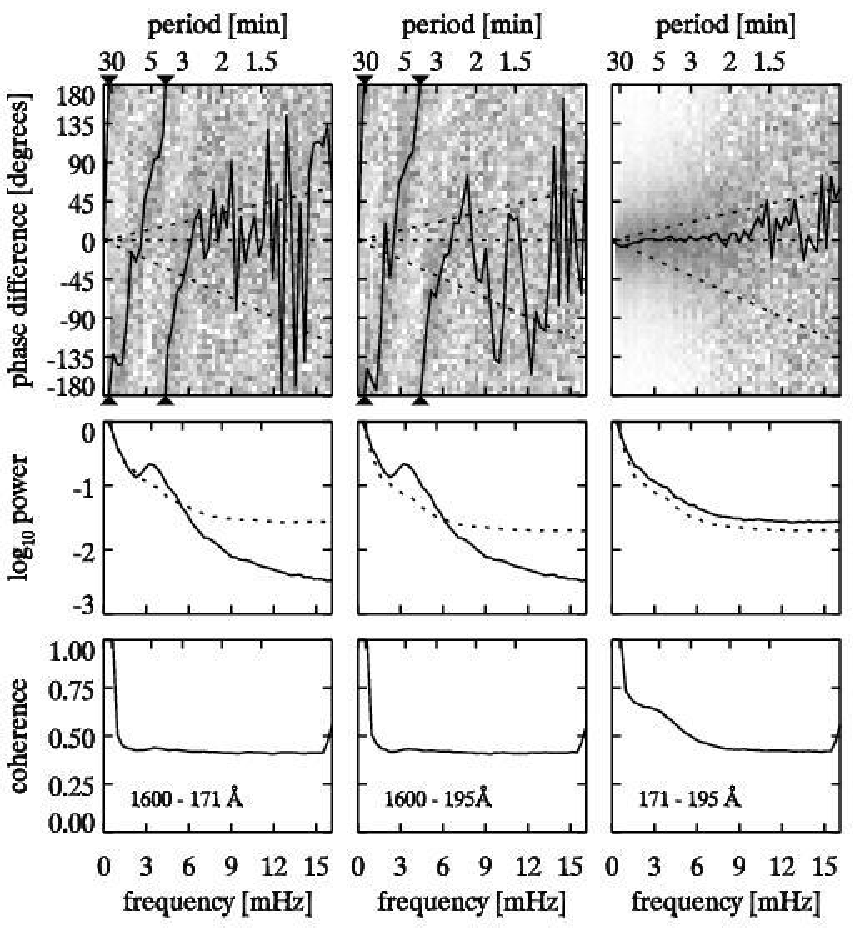}{fig:confuse1998}{%
Phase-difference and power spectra for mossy plage in the 1998 July~1 sequence.
The display and curve coding are the same as in Fig.~\ref{fig:confusomaximus}, except for the absence of non-mossy plage.}

\figureone{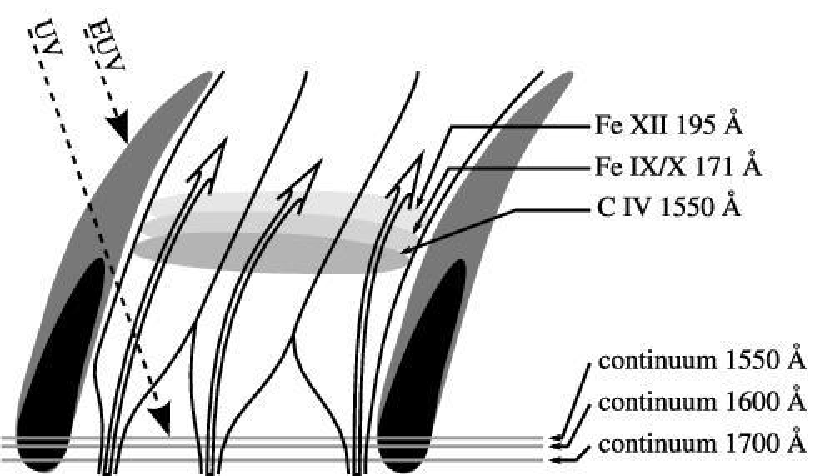}{fig:cartoon}{%
Cartoon of mossy plage.
A signal travels upward along strong field (arrows), causing modulation in all passbands.
Fibrils intermittently prevent \FeXII, \FeIXX, and \CIV\ radiation from reaching the EUV observer, thus acting as a shutter.
The shutter produces a zero phase difference between the \FeXII, \FeIXX, and \CIV\ emission, whereas the upward traveling disturbances produce a phase difference that increases linearly with frequency.
}

\figureone{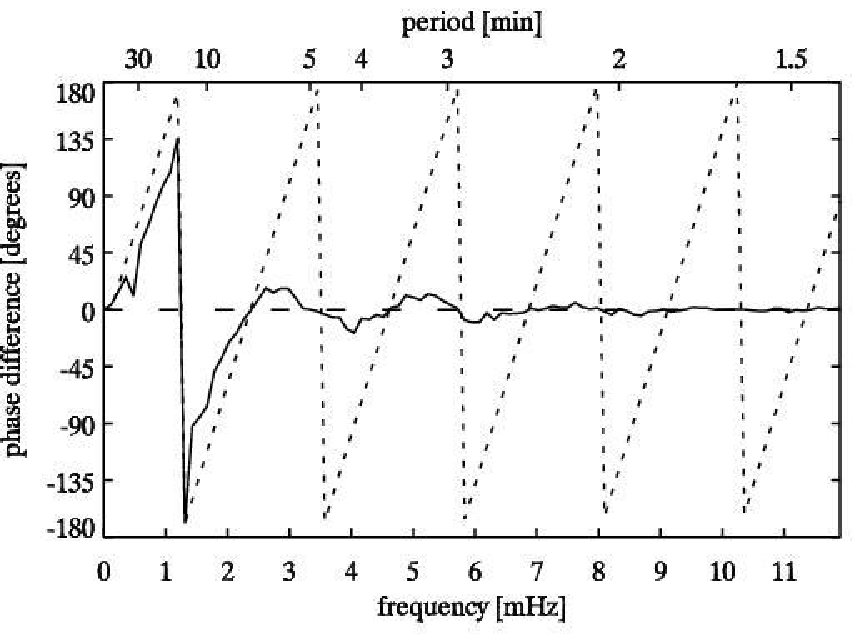}{fig:phasesim}{%
Simulated phase differences.
Solid curve: toy model with ``fibril shutter'' installed.
Dotted curve: toy model without shutter.
}

\end{document}